# Superconductivity of bulk $CaC_6$

N. Emery[1], C. Hérold[1], M. d'Astuto[2], V. Garcia[3], Ch. Bellin[2], J.F. Marêché[1], P. Lagrange[1] and G. Loupias[2]

1- Laboratoire de Chimie du Solide Minéral – UMR 7555, Université Henri Poincaré Nancy I, B.P. 239, 54506 Vandœuvre-lès-Nancy Cedex - France,

2- Institut de Minéralogie et de Physique des Milieux Condensés, IMPCM-UMR 7590, Université Pierre et Marie Curie (Paris 6), case 115, 4 place Jussieu F75252 Paris-Cedex 05,

3- Institut des Nano-Sciences de Paris, INSP-UMR 7588, Université Pierre et Marie Curie (Paris 6), 4 place Jussieu 75252 Paris-Cedex 05, France

(Dated 3 june 2005)

We have obtained bulk samples of the graphite intercalation compound, $CaC_6$, by a novel method of synthesis from highly oriented pyrolytic graphite. The crystal structure has been completely determined showing that it is the only member of the $MC_6$, metal-graphite compounds, which has rhombohedral symmetry. We have clearly shown the occurrence of superconductivity in the bulk sample at 11.5K, using magnetization measurements.

Graphite intercalation compounds (GICs) are synthesized by inserting foreign atoms or molecules between the hexagonal two-dimensional sheets of graphite, leading to ordered structures. Since graphite is a semimetal, the electrons accepted or donated by the intercalant modify the electronic properties of graphite, resulting in a metallic behavior in the final material. This process leads to a series of compounds with regular stacking of *n* graphite layers between two successive intercalant planes; *n* is referred to as the stage of the compound. Superconductivity in the first stage potassium-graphite donor compound $KC_8$, was reported years ago with a critical temperature, $T_c$, of 0.14K [1]. It is interesting that neither of the constituents are superconducting by themselves. Superconductivity in GIC's was subject of extensive studies during the late 80's and early 90's, until the discovery of higher $T_c$ in $C_{60}$ compounds [2]. In particular, several first stage metal-graphite compounds were investigated in order to test the idea that $T_c$ will increase with increasing metal concentration which is accompanied by a larger electronic charge transfer to the graphene planes [3]. However it is a fact that $LiC_6$ is not a superconductor even though the charge transferred to graphene layer [4] is shown to be larger than in $KC_8$, as expected from its stoichiometry. Nevertheless, high-pressure syntheses, leading to metastable compounds with higher metal concentration, were successful in leading to superconductivity with $T_c$ of 1.9K for $LiC_2$ [5], and 5K for $NaC_2$ [6] and $KC_3$ [7]. As a consequence, graphite has been demonstrated to be a convenient host for producing various two-dimensional ordered systems with interesting and promising electronic properties. These are not only related to electron charge transfer but also to other parameters such as noticeable changes in graphite layer spacing due to different intercalated species. It is also possible to intercalate three or even more layers of metals between to adjacent graphitic planes, leading, for example, to $KTl_{1.5}C_4$ compound, with 1210 pm interlayer distance [8], exhibiting superconductivity at 2.7K [9], up to now the highest among all the graphite intercalation compounds synthesized at ambient pressure.

Very recently, $T_c$ of 6.5K was discovered in $YbC_6$ as well as the suggestion of superconductivity at as high as 11.5K in $CaC_6$, both of them synthesized at ambient pressure [10]. However, in case of CaC6, the intercalation was limited to surface layers and as a consequence the transition temperature cannot be clearly determined since the Meissner effect measurement does not shows a clear discontinuity but only a smooth change of magnetization with temperature. In addition, the lack of saturation of diamagnetism and the very small value of diamagnetic moment are attributed by Waller *et al.* [10] to "reduced sample quality". It has been well known that the intercalation of alkaline earth metals into graphite is much more difficult than for any of alkali metals. This is particularly true in the case of calcium. Indeed, the reaction between calcium vapor and pyrolytic graphite, as described in ref. 11, leads only to an extremely superficial intercalation. For this reason, even if the distance between graphene planes reached the value of 450 pm after intercalation, as compared to 335 pm in pristine graphite, bulk $CaC_6$ was not obtained and as a result, it was not possible to determine the structure of the first stage calcium-graphite compound until now. A high quality bulk sample is needed for both accurate structure determination, and bulk superconductivity measurements.

We have succeeded in synthesizing bulk $CaC_6$, from highly oriented pyrolytic graphite [12]. The reaction is carried out for ten days between a pyrolytic graphite platelet and a molten lithium-calcium alloy at around 350°C, under very pure argon atmosphere. The reactive alloy has to be very rich in lithium, with a composition between 70 and 80 atomic percent of Li. Despite such a low calcium concentration, no lithium is present in the final reaction product and calcium alone is intercalated into graphite.

The resulting bulk sample of $CaC_6$ permitted us to carry out the first structural study of this compound [12]. The characteristic texture of pyrolytic graphite used as the precursor in the intercalation process resulted in observing the *00l* reflections in the X-ray diffraction diagram (Fig. 1). At the same time, there is an absence of *00l* reflections due to graphite or to any other GICs. The shortest distance between two graphene sheets, i.e. the interlayer distance $d_{id}$, of this pure phase is obtained from the position of the reflection peak to be equal to 452.4 pm. The stoichiometry of CaC6 was confirmed using the X-ray *00l* reflection intensities as well as by means of nuclear microprobe [13]. This stoichiometry [14] corresponds to an AA stacking of the successive graphene planes, i.e., all the carbon atoms in two successive layers are located directly on top of each other. Intercalant metal atoms in each layer are located in one out of three prismatic hexagonal sites denoted by α, β and γ. As a result, three different **c**-axis stackings have to be thus considered [15]:

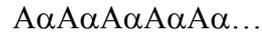
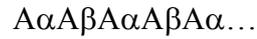
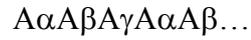

AαAαAαAαAα…

AαAβAαAβAα…

AαAβAγAαAβ…

In the first case, the c parameter of the unit cell is obviously identical to the interlayer distance ($c = d_i$). For the second stacking: $c = 2\,d_i$, and for the third one: $c = 3\,d_i$. The first and the second stackings lead to hexagonal crystal symmetry while the third corresponds to rhombohedral symmetry.

For CaC6, Waller et al [10] suggest an AαAβ stacking but our experimental results clearly show it to be AαAβAγ. Indeed, detailed analysis of *00l* and *hkl* reflections give a value of 1357.2 pm for the c parameter ($c = 3\,d_i$) of the first stage compound. In addition, this assumption is supported by the lack of reflections which obey the general diffraction conditions: *-h + k + l = 3n*, characteristic of the latter **c**-axis stacking. Clearly with such a stacking, the symmetry of crystal is rhombohedral. On the other hand, if we consider the projection the intercalant planes on an adjacent graphene layer, ignoring the dilation of the graphene plane due to electron transfer from calcium atoms, the MC6 formula leads to a characteristic 2D hexagonal unit cell, with *a* parameter equal to $a_G\sqrt{3}$ (with graphite parameter $a_G$ = 246 pm). Our value, 433.3 pm, is very close to $a_G\sqrt{3}$. Therefore, we conclude that the crystal structure of CaC6 has been conclusively determined.

We have shown that among the various MC6 graphite intercalation compounds, CaC6 is the only one to possess rhombohedral instead of hexagonal symmetry. The elemental unit cell is a rhombohedron belonging to the $R\overline{3}m$ space group with the following parameters: $a_R$ = 517 pm and α = 49.55°. It contains one calcium atom in site (0  0  0) and six carbon atoms in sites (1/6  5/6  1/2) and is centro-symmetrical and belonging to the holoedry of rhombohedral system. However, it is often preferable to describe such a structure using a hexagonal unit cell, whose size is three times that of the rhombohedral cell (Fig. 2).

In addition, from the measured dilatation of carbon-carbon bond compared to pure graphite, we estimate the electron transfer between calcium and graphene plane, using the empirical Pietronero-Strässler rule [16]. It reaches the high value of 0.103 electrons per carbon atom and appears to be the highest charge transfer evaluated among all the $MC_6$ compounds.

In order to study the details of bulk superconductivity in $CaC_6$, its magnetization was measured as a function of temperature and magnetic field. Two different samples were used to perform two sets of measurements using a Quantum Design MPMS5 SQUID magnetometer. Since the **c** axes of the all crystallites forming the highly oriented graphite are parallel to each others, we are able to define the **c** axis direction of the $CaC_6$ sample, while in the perpendicular direction the material is disordered, leading to an average of **a** and **b** directions, denoted as *ab*. Because of their chemical reactivity, the samples had to be kept in a closed cell, under helium atmosphere. The first sample, with a roughly rectangular shape of about 3,5mm x 3mm and thickness of 0.3 mm was measured with the magnetic field in the *ab* plane. The second sample was composed of 3 superposed platelets of about 8 $mm^2$ surface and 0.2 mm thickness each, was measured with the magnetic field along the **c** axis.

Magnetization as a function of temperature was measured for each sample at two different values of externally applied magnetic field. The samples were initially cooled to 2K with the magnetic field switched off and subsequently the magnetization was measured with magnetic fields of 5 and 50 Oe, increasing the temperature up to 40K. This is referred to as "zero field cooling" experiment (ZFC). The same measurement was also carried by cooling the sample with the magnetic field applied, referred to as "field cooling" experiment (FC). Figure 3 shows the results for both ZFC and FC measurements at 5 Oe and at intervals of 0.2K up to 20K.

The ZFC results exhibit a sharp drop of the magnetization below 11.5 K, with 2/3 of the decrease occurring within 0.2K. The strong diamagnetic signal shows a complete saturation below the superconducting transition temperature, indicating a true Meissner effect. This was not observed by Weller *et al.*[10]. The midpoint for the ZFC curve occurs at 11.1K with a total transition width $\Delta T$ of 0.5 K, using temperatures corresponding to 90% and 10% of the diamagnetic signal. Moreover a series of magnetization measurements versus the field were carried out at given temperatures around $T_C$, by steps of 0.1K, allowing us to determine precisely that the diamagnetic signal completely disappears at $T_c^{onset}$ = (11.46±0.04)K. The transition temperature appears as the highest transition temperature observed among all the graphite intercalated compounds.

The FC results show a similarly sharp drop with the same onset of 11.5K. However the diamagnetic signal is strongly reduced compared to the ZFC. The superconducting volume fraction inferred from the diamagnetic signal is of 60%, with an error bar of 10%, obtained taking into account both demagnetization corrections and sample volume estimations. This value confirms the bulk intercalation and not only surface intercalation. Similar results were obtained for measurements at 50 Oe in the same direction and for 50 Oe and 5 Oe with the magnetic field applied along the **c** axis.

Figure 4 shows the temperature dependence of the critical fields $H_{c1}$ and $H_{c2}$ obtained from the magnetization versus field, M(H), measurements at a given temperature (the 6K data are shown in the inset) and with the field oriented along the **c**-axis. The material appears to be a type-II superconductor. Note that the $H_{c2}$ value at 2 K is a lower bound, since the measurements were performed only up to 2500 Oe and a very small diamagnetic signal (about $4 \times 10^{-4}$ emu) was still present. The overall T-dependence of $H_{c1}$ and $H_{c2}$ is compatible with the dependence derived by Helfand and Werthamer [17] within the BCS theory. In particular, in the region close to $T_C$, the dependence is roughly linear, in agreement with a Ginzburg-Landau picture, with an upturn at T > 9K, possibly associated with fluctuation effects. The extrapolation of this tail is consistent with the measured $T_c^{onset}$ = 11.5K. From similar measurements, with the field in the *ab* plane, we estimate a higher upper critical field $H_{c2}$ of about 7000 Oe, indicating a sizeable anisotropy of the critical field, as already observed in $C_6Yb$ [10]. From the linear part of our measurements, we can evaluate the coherence lengths both in the *ab*-plane and along the **c** axis respectively to be $\xi_{ab,0}$ = 350 Å and $\xi_{c,0}$ = 130 Å. This implies that the superconducting state of this material is anisotropic but clearly three-dimensional in spite of the layered structure.

In order to find GICs exhibiting higher superconducting transition temperature than $KC_8$, two main paths have been previously followed: increasing the interlayer distance, $d_i$ [9] and increasing the metal concentration [3]. The goal of both these two approaches was to obtain high and well-localised electronic density in the graphitic planes. In the present study of $CaC_6$, a transition temperature of more than two times larger than $KC_3$ [7] was obtained. We notice that the measured interlayer distance of 452 pm in $CaC_6$ is smaller than 535 pm in $KC_8$, making the first approach unlikely to succeed. This is in contrast to the situation for the alkaline $C_{60}$ compounds, another series of carbon-based intercalated compounds, where the alkali atom plays the role of spacers between the $C_{60}$ molecules, and the superconducting temperature rises from 19.4K for $K_3C_{60}$ [2] to 32K for $CsRb_2C_{60}$ [18] with an increase in the intercalant atom diameter. Similarly the maximum charge transfer, expected from $CaC_6$ stoichiometry, is equal to two electrons given by each Ca atom, i.e. one third of electron accepted by each carbon atom, the same as for $KC_3$ and lower than for $LiC_2$, one half electron per carbon atom, with a superconducting temperature as low as 1.9K. Moreover, the estimated charge transfer of 0.1 electron per carbon atom for $CaC_6$ is three times less that the largest possible value. As a result, the simple increase of charge transfer from the intercalant to graphite sheet appears to play a questionable role in this process. Recent calculations by Calandra and Mauri [19] indicate that the existence of substantial amount of charge left on the metal plays a major role in existence of superconductivity and magnitude of the transition temperature. In addition, the authors have notice that the Ca rhombohedron in $CaC_6$ is close to the elementary cell of the fcc Ca pure metal, leading to a sizeable 3D character of metallic bands in the intercalated compound. This result is supported by the low anisotropy of the coherence lengths $\xi_{ab}$ and $\xi_c$, obtained from our measurements.

In conclusion, we have unambiguously demonstrated the existence of bulk superconductivity in $CaC_6$ both by the sharpness of the measured magnetization versus temperature and by the saturation of diamagnetic signal. Our novel and efficient method for preparation of high quality bulk $CaC_6$ played a crucial role in accurate measurement of the superconducting temperature as well as the width of the transition. We have also shown that among the various $MC_6$ graphite intercalation compounds, $CaC_6$ is the only one with rhombohedral instead of the hexagonal symmetry as postulated elsewhere [10].

The three families of carbon-based materials, graphite intercalated compounds [1], doped fullerines [18] and doped diamond [20], have all exhibited superconductivity. However, the present case of $CaC_6$ is particularly significant since it represents an increase of Tc by an order of magnitude, at ambient pressure, over that of $KC_8$, discovered 40 years ago. This has led to renew interest in taking advantage of the wide variety of intercalant species and structures in carbon-based materials. Furthermore, we have shown that the simple models existing until now are insufficient to predict the magnitude of the superconducting transition in these materials.

We acknowledge F. Mauri for interesting suggestions and fruitful discussions. We also thank M. Calandra, A. Gauzzi, S. Rabii, A. Shukla, and M. Marangolo for helpful discussions. Thanks are due to Cyril Train for providing access to SQUID.

**Figure captions:**

Figure 1: *00l* diffraction pattern for $CaC_6$, using Mo $K\alpha_1$ X-rays.

Figure 2: (color online) Rhombohedral unit cell and the corresponding hexagonal cell of $CaC_6$.

Figure 3: (color online) Magnetization of $CaC_6$ as a function of temperature for an external magnetic field of 5 Oe under zero field cooling (ZFC) and field cooling (FC) conditions.

Figure 4: (color online) Superconducting phase diagram of $CaC_6$ for the magnetic field along the **c** axis from M(H) measurements at given T. The 6K measurement is shown in the inset.

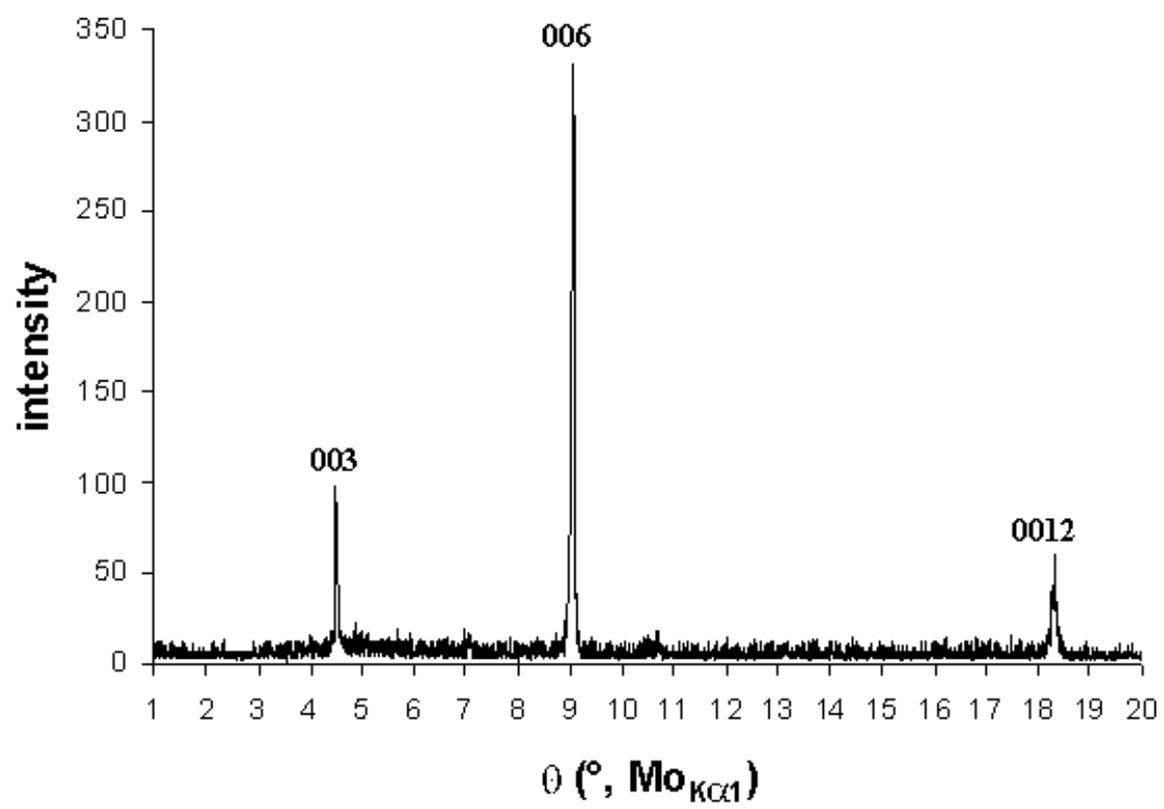

Figure 1

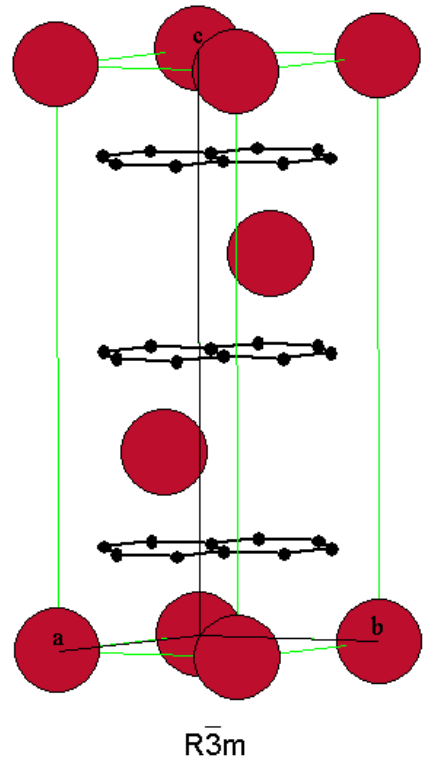

Figure 2

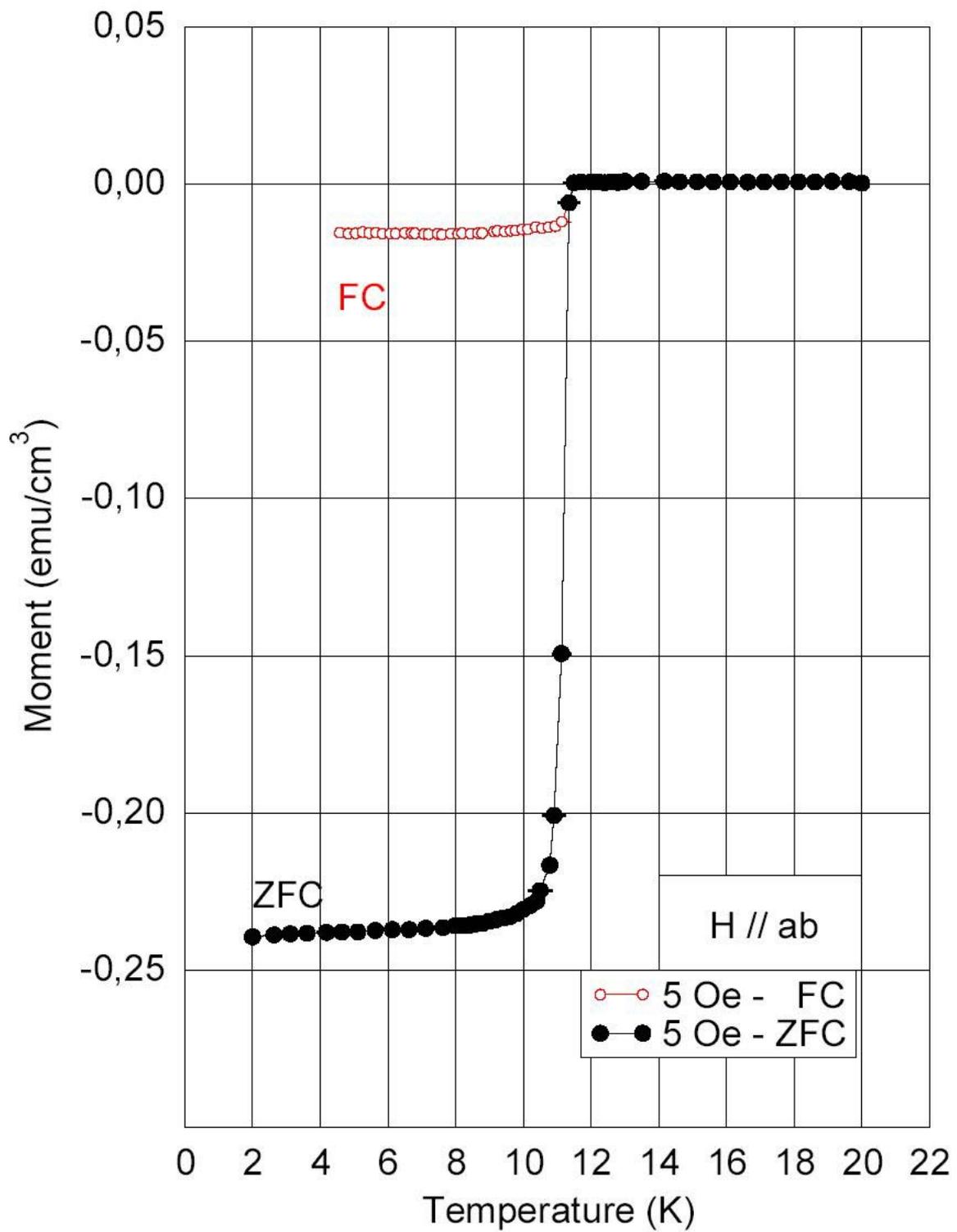

Figure 3

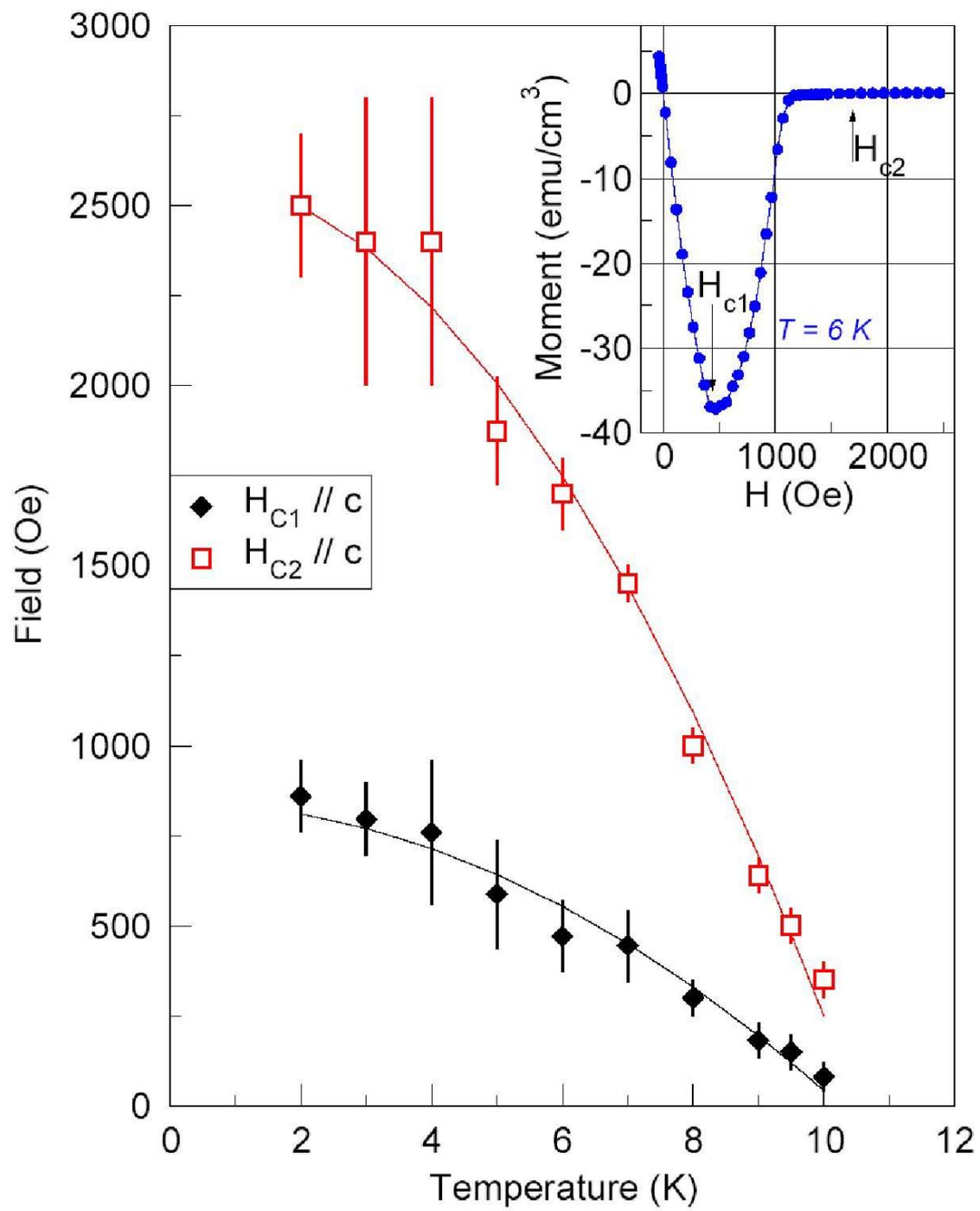

Figure 4